# A very young and fast rotating shell star discovered in the eclipsing binary ZTF J200347.63+394429.8

Norbert Hauck

Bundesdeutsche Arbeitsgemeinschaft für Veränderliche Sterne e.V. (BAV), Munsterdamm 90, 12169 Berlin, Germany; hnhauck@yahoo.com

**Abstract:** *A photometric study in combination with existing stellar models has revealed details of this eclipsing post-mass-transfer binary. The shell star has an equatorial/polar radius of ~2.60/1.90 Rsun at an equatorial rotational velocity of ~430 km s$^{-1}$, an effective mean temperature $T_{eff}$ of ~13300 K and a mass of ~3.42 Msun. This former accretor star is surrounded by a large decretion disk of ~47 Rsun. The secondary star is a helium white dwarf precursor with a radius of 0.98 Rsun, a $T_{eff}$ of ~17100 K and a mass of 0.29 Msun. The parameters of this former donor star indicate an age of the binary system of only ~1.8 Myr after the end of mass transfer. The results fit to a sub-solar metallicity of Z = 0.007.*

The eclipsing binary ZTF J200347.63+394429.8 (GAIA DR3 2074020676273655296; 2MASS J20034764+3944298) has been mentioned in the Zwicky Transient Facility (ZTF) catalog of periodic variable stars of Chen et al. [1], including an orbital period of ~45 days. It is a stellar object of ~13.8 *V*mag brightness and located in the Cygnus constellation. No spectral classification has been found in the literature.

A first light curve has been produced from existing ZTF data points (see Fig. 1). It does show a remarkable similarity with the peculiar one already encountered in the eclipsing shell star binary V658 Car (see Fig. 1-3 of Hauck [2]). New photometric data in the passbands *UBV* have been obtained with remotely controlled 17-, 20- and 24-inch telescopes in Utah and California, USA. The light curves have been modeled with the *Binary Maker 3* (BM 3) software. The total brightness of the binary system in *UBV* (see Table 1) has been taken from GAIA DR3, part 6, synthetic photometry (CDS cat. I/360/syntphot).

The **primary minimum** (~14.6 hours) of our best fit solution is nearly a total eclipse, i.e. missed by only 0.03° of orbital inclination, since the separation of the stars is a bit larger in this position of the somewhat eccentric orbit than at the **secondary minimum** (~13.1 hours) showing an annular eclipse of ~2.1 hours (see Fig. 2 - 4). The dominant, deep **disk eclipse** lasting 9.3 days from phase 0.9055 to 0.1136 (see Fig. 1) indicates a slightly eccentric position w.r.t. the primary star in its center. Moreover, adopting a Roche lobe filling disk, this phase difference indicates a mass ratio of 11.80 ± 0.035 in our binary solution.

The primary eclipse being exclusively attributed to the loss of light of the smaller companion star allows the calculation of its **temperature** $T_{eff}$ as well as the interstellar extinction from our *UBV*-photometry in phase 0 and at maximum light. Adopting a normal interstellar extinction ($R_v$ = 3.16) gives the coefficients $A_U$ = 1.584$A_V$ and $A_B$ = 1.317$A_V$ according to Table 3 of Wang & Chen [3]. From Table 5 of Pecaut & Mamajek [4] containing the $T_{eff}$-dependent values of *U–B* and *B–V* we finally get a $T_{eff}$ of 17140 ± 570 K for the secondary star and an interstellar extinction $A_V$ of 1.00 ± 0.02 mag.

Fitting the light curve to our *V*-band data points of the secondary minimum (see Fig. 3) (plus an artificial total eclipse) results in an apparent mean $T_{eff}$ of 8370 K for the primary shell star in our equatorial view (behind half of its *densified* decretion disk) at a rotation rate factor of 146.3 times per orbital cycle. Towards maximum light there is a visible brightening in the light curve (see Fig. 1) corresponding to an increased mean $T_{eff}$ of 8680 K for the shell star behind half of its *undensified* disk ($A_V$ half disk = 0.887 mag), and a mean $T_{eff}$ of 11590 K without any disk in our equatorial view. The dimming effect of the entire disk of 78.5, 78.0 and 80.4 % of the flux in *V*, *B* and *U*, respectively, has been determined from observations in phase 0.0197 shortly after primary stellar eclipse. For the mean $T_{eff}$ of the total surface area of this rotating star we get 13260 K (by comparison with a sphere of the same surface area and luminosity at 550 nm). The therefrom calculated theoretical $T_{eff}$ of the non-rotating star is 14750 K (with the help of Fig. 4 and 5 of Ekström et al. [5]).

Further **parameters** of our binary components (see Table 2) have been derived from the above adopted mass ratio of 11.80. For the secondary star a mass of 0.290 $M_\odot$, a radius of 0.980 $R_\odot$ and a bolometric luminosity of 74.34 $L_\odot$ then fit into the stellar models for helium white dwarf precursors, i.e. Fig. 1 of Driebe et al. [6] combined with Fig. 9 of Istrate et al. [7]. For the primary star we now get a mass of 3.42 $M_\odot$, and a theoretical radius of 1.927 $R_\odot$ with a bolometric luminosity of 157.6 $L_\odot$ for the *non-rotating* star, which nicely fit at a half-solar metallicity Z of 0.007 *between* the stellar models of Georgy et al. [8] (Z 0.006, 3.38 $M_\odot$, 38 Myr after zero-age main sequence (ZAMS)) and Schaerer et al. [9] (Z 0.008, 3.46 $M_\odot$, on the ZAMS). The half-way approach of our rejuvenated star towards the ZAMS, after having achieved thermal equilibrium again, is in line with the expected behavior (see C' in Fig. 3.7(a), Eggleton [10]). Finally, the successful *simultaneous* fit of both stars into their stellar models confirms the adopted mass ratio.

Our best fit light curve solution includes an amount of decretion disk light, i.e. at a wavelength of 550 nm, 25.2 % at first resp. fourth contact of the stellar secondary minimum and 28.3 % at maximum light. Thereby, the distance of the shell star is in line with the one that has been calculated for the pre-He white dwarf, i.e. ≈ 2938 pc.

The results fit to an **evolutionary history** of a post-mass-transfer binary in which an initially more massive star of 2.19 $M_\odot$ has transferred (by an assumed conservative Roche lobe overflow) his hydrogen shell to a companion star of 1.52 $M_\odot$, and now, about 1.8 Myr after the end of mass transfer, has become an unusually hot and massive contracting helium white dwarf precursor. Helium stars with masses of < ~0.3 $M_\odot$ do not burn He, but become degenerate (Eggleton [10], section 2.5). This mass transfer has also caused the very rapid rotation of the accretor, which now is a shell star being surrounded by an equatorial decretion disk having a mean radius of ~47 $R_\odot$.

Apart from its remarkable flattening ratio of 1.37 ± 0.04 (critical limit = 1.5) and a rotational velocity ratio W = $v_{rot}/v_{orb}$ of 0.86 ± 0.03 (critical limit = 1.0) our shell star has a calculated **rotational velocity** of 432 ± 9 km s$^{-1}$ at the equator, which is surprising in the light of other young post-mass-transfer binary systems. Rivinius et al. [11] have recently given rotational velocities of up to ~ 60 km s$^{-1}$ for similar newborn Be star systems (see their table 4). Our new shell star, however, appears to be one of the fastest rotators of B-type stars found to date.

A search in the literature for rotational velocities significantly above 450 km s$^{-1}$ in normal, i.e. non-degenerate stars, has delivered only three convincing findings: LAMOST J040643.69+542347.8 (Li [12]), VFTS 285 & VFTS 102 (Shepard et al. [13]) with projected rotational velocities vsin $i$ of 540 ± 29, 610 ± 41 and 649 ± 52 km s$^{-1}$, respectively. They all belong to the more massive O-type stars. Moreover, they have either a suspected metallicity of the Large Magellanic Cloud (LMC) in case of the Milky Way (MW) LAMOST star, or they are members of the LMC. Lower metallicity implies faster stellar rotation and less incidence of stellar pulsations (Rivinius et.al. [14], see their Table 3 as well as Fig. 18 and the interesting completion by our primary star). Therefore, the sub-solar (~LMC) metallicity of Z = 0.007 surprisingly well fitting to our Milky Way shell star appears to be an explanation for its unusually fast rotation.

**Acknowledgements:**

This publication is based on observations obtained with the Samuel Oschin 48-inch Telescope at the Palomar Observatory as part of the Zwicky Transient Facility project. This research has also made use of the Simbad and VizieR databases operated at the Centre de Données astronomique Strasbourg (CDS), France.

**Table 1 : Parameters of the binary system ZTF J200347.63+394429.8**

| | | |
|---|---|---|
| Epoch [HJD] | 2458657.955(14) | mid primary minimum |
| Orbital period [days] | 44.685(1) | ( ) = errors in last decimals |
| Apparent $V$ magnitude | 13.820(2) | from GAIA DR3, part 6 |
| Apparent $B$ magnitude | 13.994(3) | from GAIA DR3, part 6 |
| Apparent $U$ magnitude | 13.743(11) | from GAIA DR3, part 6 |
| Eclipse duration [hours] | 14.6 | primary minimum |
| Eclipse duration [hours] | 13.1 | secondary minimum |
| Eclipse duration [days] | 9.3 | disk eclipse |
| Eccentricity $e$ | 0.075 | |
| Longitude of periastron $\omega$ [deg] | 61 | for primary star's orbit |
| Orbital inclination $i$ [deg] | 89.37 ± 0.08 | |
| Semi-major axis $a$ [$R_\odot$] | 81.97 ± 0.38 | for $R_\odot$ = 696342 km |
| Distance [pc] | 2938 ± 79 | |
| Extinction $A_V$ [mag] | 1.00 ± 0.02 | interstellar |

**Table 2 : Parameters of the components of ZTF J200347.63+394429.8**

| Parameter | primary star | secondary star | disk |
|---|---|---|---|
| Radius (mean) [$R_\odot$] | 2.37 ± 0.14 | 0.980 ± 0.013 | 47 |
| Radius (pole/equator) [$R_\odot$] | 1.90 / 2.60 | | |
| Rotational velocity [km s$^{-1}$] | 432 ± 9 | | |
| Temperature mean $T_{eff}$ [K] | 13260 ± 440 | 17140 ± 570 | |
| $V$-flux fraction at max. light | 0.3798 | 0.3369 | 0.2833 |
| $B$-flux fraction at max. light | | 0.3486 | |
| $U$-flux fraction at max. light | | 0.4043 | |
| Apparent $V$ magnitude | 14.871 | 15.001 | |
| Apparent $B$ magnitude | | 15.138 | |
| Apparent $U$ magnitude | | 14.726 | |
| Luminosity (bolometric) [$L_\odot$] | 146.45 | 74.34 | |
| Mass [$M_\odot$] | 3.42 ± 0.05 | 0.290 ± 0.004 | |

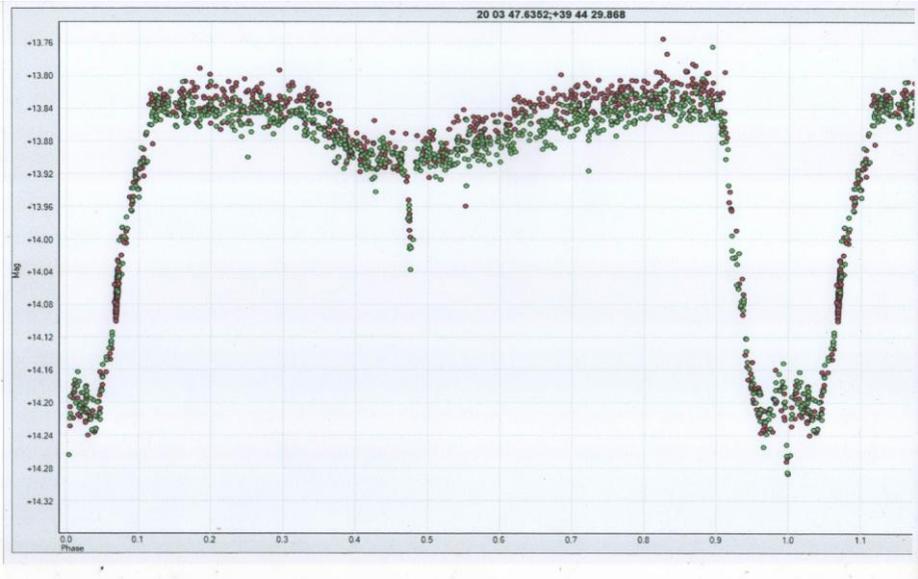

**Fig. 1:** The light curve of our eclipsing shell star binary from ZTF data points in the passbands *g* (496 nm) in green color and *r* (621 nm) in red color, folded with our orbital period of 44.685 days. The light curve shape is dominated by dimming effects of the decretion disk, but the stellar eclipses in phase 0 and 0.4763 are still visible.

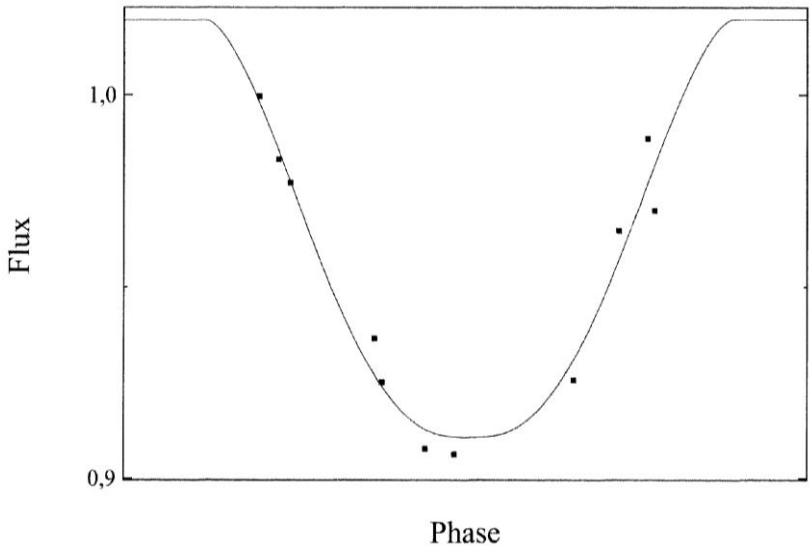

**Fig. 2:** Light curve fit (σ 8 mmag) to ZTF *g*-data points at primary minimum (phase 0).

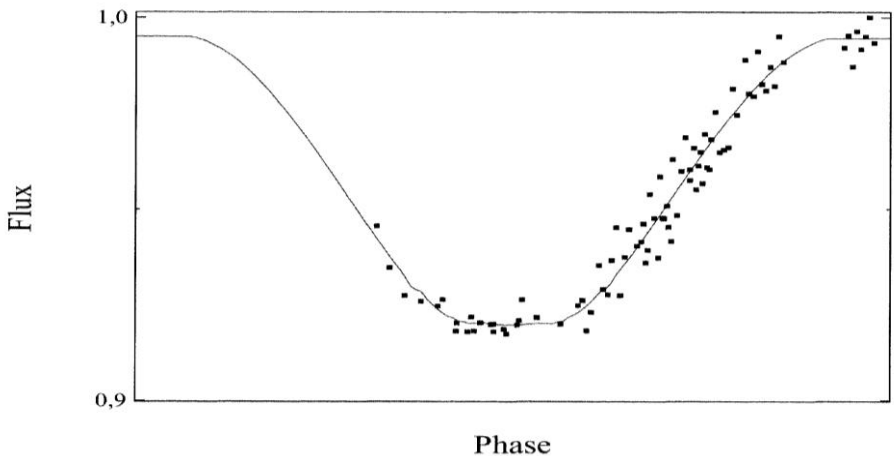

**Fig. 3:** Light curve fit ($\sigma_{FIT}$ = 5.5 mmag) to our data points in passband *V* at the secondary minimum in phase 0.4702 - 0.4824 (annular in phase 0.4753 - 0.4773).

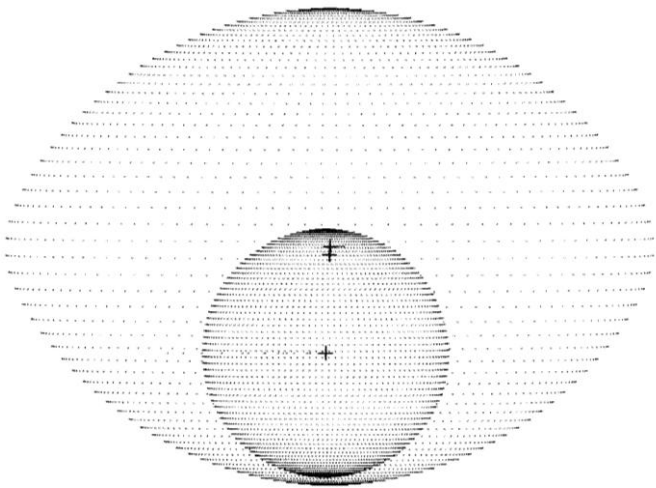

**Fig. 4:** Our equatorial view of the very fast rotating shell star being eclipsed by the pre-He white dwarf at mid secondary minimum (phase 0.4763).